\documentclass[
aps,
prc,
reprint,
superscriptaddress,
nofootinbib,
showkeys
]{revtex4-2}

\usepackage[utf8]{inputenc}
\usepackage[T1]{fontenc}
\usepackage{amsmath,amssymb,bm}
\usepackage{graphicx}
\usepackage{xcolor}
\usepackage{booktabs}
\usepackage{array}
\usepackage{multirow}
\usepackage{dcolumn}
\usepackage{mathrsfs}
\usepackage{tabularx}
\usepackage{array}
\usepackage{microtype}

\definecolor{linkblue}{RGB}{0,70,150}
\definecolor{citered}{RGB}{150,35,35}
\definecolor{urlpurple}{RGB}{100,40,150}

\usepackage[
colorlinks=true,
linkcolor=linkblue,
citecolor=citered,
urlcolor=urlpurple
]{hyperref}

\IfFileExists{orcidlink.sty}{
  \usepackage{orcidlink}
}{
  \newcommand{\orcidlink}[1]{\href{https://orcid.org/#1}{\textsuperscript{\scriptsize ORCID}}}
}

\newcommand{\mucf}{\ensuremath{\mu{\rm CF}}}
\newcommand{\dtmu}{\ensuremath{dt\mu}}

\newcommand{\alphamu}{\ensuremath{\alpha\mu}}
\newcommand{\Lmu}{\ensuremath{\mathcal{L}_\mu}}
\newcommand{\Lamc}{\ensuremath{\Lambda_c}}
\newcommand{\taumu}{\ensuremath{\tau_\mu}}

\newcommand{\weff}{\ensuremath{\omega_S^{\rm eff}}}
\newcommand{\wcrit}{\ensuremath{\omega_{S,{\rm crit}}^{\rm eff}}}
\newcommand{\Ecost}{\ensuremath{E_\mu^{\rm cost}}}
\newcommand{\Euse}{\ensuremath{E_{\rm use}}}
\newcommand{\Nfus}{\ensuremath{N_{\rm fus,\mu}}}
\newcommand{\Gmu}{\ensuremath{G_\mu}}

\newcommand{\FigPlaceholder}[2]{
\begin{center}
\fbox{
\begin{minipage}[c][#2][c]{#1}
\centering
{\large Figure placeholder}\\[0.15cm]
\texttt{fig/fig\_main\_lawson\_two\_panel\_pub\_v6}\\[0.10cm]
Replace this box by the production-quality PDF or PNG.
\end{minipage}
}
\end{center}
}

\newcommand{\MainLawsonFigure}{
\IfFileExists{fig/fig_main_lawson_two_panel_pub_v6.pdf}{
\includegraphics[width=0.98\textwidth]{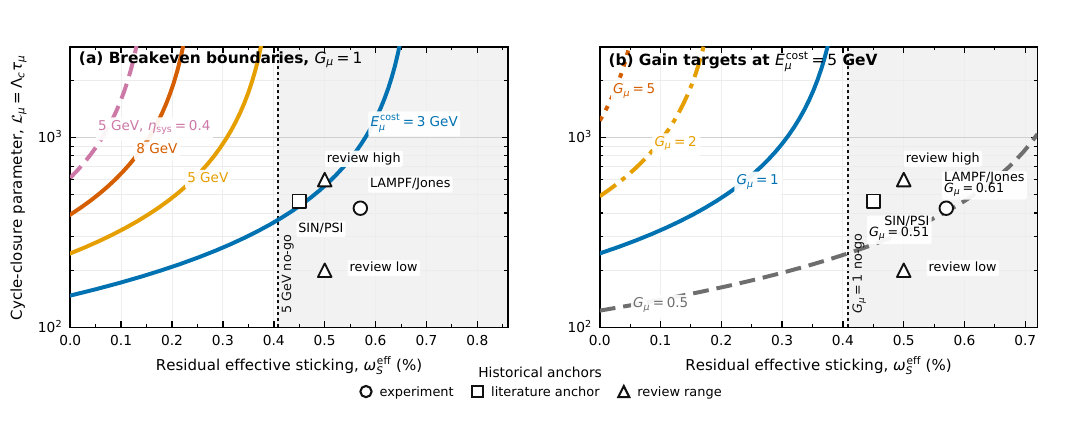}
}{
\IfFileExists{fig/fig_main_lawson_two_panel_pub_v6.png}{
\includegraphics[width=0.98\textwidth]{fig/fig_main_lawson_two_panel_pub_v6.png}
}{
\FigPlaceholder{0.92\textwidth}{4.2cm}
}
}
}

\begin{document}

\title{
A Lawson-inspired Cycle-Closure Criterion for Deuterium--Tritium Muon-Catalyzed Fusion
}

\author{Wei Kou\orcidlink{0000-0002-4152-2150}}
\email{kouwei@impcas.ac.cn}
\affiliation{Institute of Modern Physics, Chinese Academy of Sciences, Lanzhou 730000, Gansu Province, China}
\affiliation{Southern Center for Nuclear Science Theory (SCNT), Institute of Modern Physics, Chinese Academy of Sciences, Huizhou 516000, Guangdong Province, China}
\affiliation{School of Nuclear Science and Technology, University of Chinese Academy of Sciences, Beijing 100049, China}
\affiliation{State Key Laboratory of Heavy Ion Science and Technology, Institute of Modern Physics, Chinese Academy of Sciences, Lanzhou 730000, Gansu Province, China}

\author{Xurong Chen}
\email{xchen@impcas.ac.cn (Corresponding author)}
\affiliation{Institute of Modern Physics, Chinese Academy of Sciences, Lanzhou 730000, Gansu Province, China}
\affiliation{Southern Center for Nuclear Science Theory (SCNT), Institute of Modern Physics, Chinese Academy of Sciences, Huizhou 516000, Guangdong Province, China}
\affiliation{School of Nuclear Science and Technology, University of Chinese Academy of Sciences, Beijing 100049, China}
\affiliation{State Key Laboratory of Heavy Ion Science and Technology, Institute of Modern Physics, Chinese Academy of Sciences, Lanzhou 730000, Gansu Province, China}



\begin{abstract}
Deuterium--tritium muon-catalyzed fusion is limited by a cycle-closure problem: a negative muon must complete enough catalytic cycles before decay or effective alpha sticking removes it from reuse. We formulate a Lawson-inspired criterion for this single-muon cycle. The effective cycle strength is defined as $\Lmu=\Lamc\taumu$, where $\Lamc$ is the effective cycle-completion rate and $\taumu$ is the muon lifetime. Together with the residual effective sticking probability $\weff$, it gives the mean fusion yield per useful muon, $\Nfus=\Lmu/(1+\weff\Lmu)$. Introducing the useful D--T cycle energy $\Euse$, the system factor $\eta_{\rm sys}$, and the effective wall-plug-equivalent cost $\Ecost$ of one delivered useful $\mu^-$, the one-muon gain is $\Gmu=(\eta_{\rm sys}\Euse/\Ecost)\Nfus$. This leads to the required cycle strength $\Lmu^{\rm req}=\Gmu N_L/(1-\weff\Gmu N_L)$, with $N_L=\Ecost/(\eta_{\rm sys}\Euse)$, and to the conditional sticking boundary $\weff<1/(\Gmu N_L)$. The criterion separates rate-limited, sticking-limited, and cost-limited regimes in the $(\weff,\Lmu)$ plane. When representative reported or inferred D--T \mucf{} anchors are projected onto this plane, they lie in a high-yield region but remain constrained by the effective-sticking boundary under conventional multi-GeV wall-plug-equivalent muon-cost accounting. The framework provides a compact diagnostic for assessing whether future improvements act mainly by increasing the effective cycle-completion rate, reducing residual sticking, or lowering the useful cost of delivered muons.
\end{abstract}

\keywords{
muon-catalyzed fusion;
Lawson-inspired criterion;
alpha sticking;
D--T fusion;
cycle closure;
muon source
}

\maketitle


\section{Introduction}

\begin{figure}[htbp]
\centering
\includegraphics[width=0.98\columnwidth]{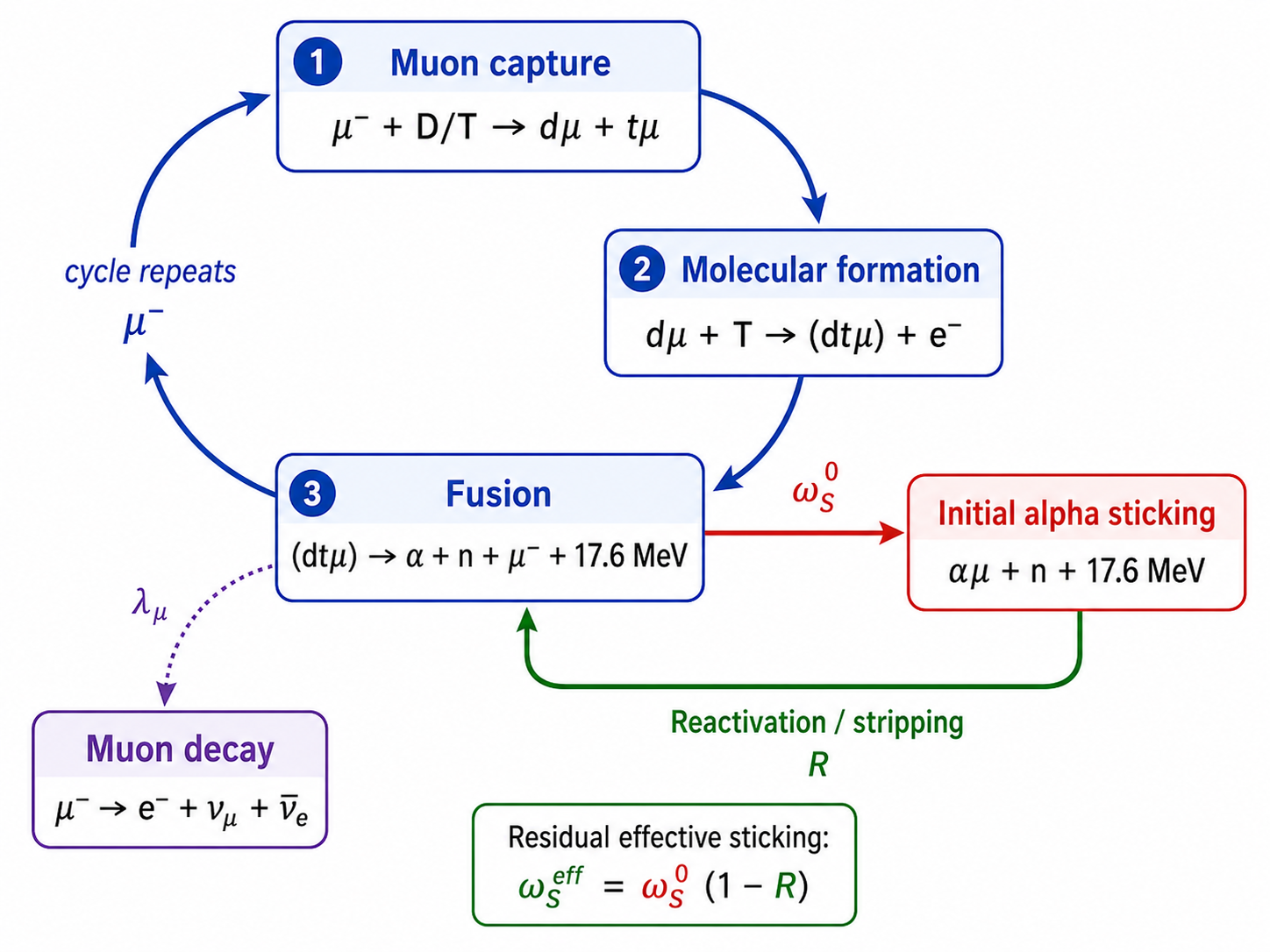}
\caption{
Schematic cycle of D--T muon-catalyzed fusion.
The main catalytic sequence consists of muon capture, $\dtmu$ molecular formation, and fusion.
After fusion, the muon may be released to continue the cycle, lost through ordinary muon decay, or trapped in an $\alphamu$ state with initial sticking probability $\omega_S^0$.
Reactivation or stripping, characterized by the probability $R$, reduces the residual effective sticking to $\omega_S^{\rm eff}=\omega_S^0(1-R)$.
}
\label{fig:mucf_main}
\end{figure}

Muon-catalyzed fusion (\mucf) provides a rare low-temperature route to nuclear fusion by replacing an electron in a hydrogen isotope with a negative muon, thereby shrinking the characteristic atomic or molecular length scale by roughly the muon-to-electron mass ratio. In a deuterium--tritium mixture, the $\dtmu$ molecule is the most favorable catalytic species: intramolecular D--T fusion releases an alpha particle, a $14.1~{\rm MeV}$ neutron, and, in most events, a muon that may initiate another cycle. The same process also contains its intrinsic loss channel, because a fraction of the emitted muons remain bound to the alpha particle as an $\alphamu$ state and are removed from the cycle unless reactivated. The basic catalytic cycle, together with the dominant loss and reactivation channels, is illustrated in Fig.~\ref{fig:mucf_main}. Since the early proposal and observation of muon-induced nuclear reactions, \mucf{} has been studied as both a few-body atomic-nuclear problem and a potential source of fusion energy or intense fusion neutrons \cite{frank1947hypothetical,Alvarez:1957un,jackson1957catalysis,ponomarev1990muon,froelich1992muon,Petitjean:1992iq}. Recent coupled-channel and $T$-matrix calculations of the $(\dtmu)_{J=v=0}$ reaction have refined the microscopic inputs, including the intramolecular fusion rate, the initial alpha-sticking probability, and the emitted-muon spectrum \cite{Kamimura:2021msf,Wu:2024uad}.

The central difficulty for energy closure is therefore not whether the $\dtmu$ molecule can fuse rapidly, but whether one useful negative muon can complete enough catalytic cycles before ordinary decay or effective alpha sticking terminates its reuse. In standard kinetic language, the fusion yield per muon is commonly estimated as $Y_f\simeq(\omega_S^{\rm eff}+\lambda_0/\lambda_c\phi)^{-1}$, where $\lambda_0=1/\tau_\mu$ is the muon decay rate, $\lambda_c$ is the effective cycle-completion rate, $\phi$ is the target density normalized to liquid hydrogen density, and $\omega_S^{\rm eff}$ is the residual effective sticking probability after reactivation. This expression correctly identifies the two microscopic bottlenecks, rate limitation and sticking loss, and it explains why measured D--T \mucf{} yields have remained of order $10^2$ fusions per muon \cite{Ackerbauer:1999,Bom:2005PAN,Kamimura:2021msf}. However, yield alone is not yet a closure criterion: it must be compared with the useful energy assigned to each D--T cycle, the efficiency of the surrounding system, and the effective wall-plug-equivalent cost of producing, transporting, moderating, and stopping a usable negative muon in the target. This missing conversion from kinetic yield to gain boundary is the gap addressed in the present work.

The purpose of the present work is to convert this kinetic information into a compact Lawson-inspired cycle-closure criterion. In the classical Lawson argument, microscopic fusion reactivity becomes useful only after it is combined with confinement and energy-balance requirements \cite{Lawson1957Criterion}. The analogy developed here is structural rather than thermodynamic: the relevant confinement variable is not a plasma energy-confinement time, but the finite lifetime during which a muon remains available for repeated catalysis. We therefore define the dimensionless cycle strength as $\Lmu=\Lamc\taumu$, where $\Lamc$ is the effective cycle-completion rate and $\taumu$ is the muon lifetime. Together with the residual effective sticking probability $\weff$, this gives the renewal yield $\Nfus=\Lmu/(1+\weff\Lmu)$. Combining this relation with a target one-muon gain $\Gmu$ and the wall-plug-equivalent cycle demand $N_L=\Ecost/(\eta_{\rm sys}\Euse)$ gives the required cycle strength $\Lmu^{\rm req}=\Gmu N_L/(1-\weff\Gmu N_L)$ and the associated sticking boundary $\weff<1/(\Gmu N_L)$. Thus the usual yield expression becomes an operational test: a system must first lie below the sticking boundary and then reach the required value of $\Lmu$.

The resulting formulation is most useful as a diagnostic map. Historical D--T \mucf{} results may be projected onto the $(\weff,\Lmu)$ plane either from reported kinetic parameters or, when a yield and a reported, inferred, or adopted effective-sticking value are available, through the inverse relation $\Lmu=Y_f/(1-\weff Y_f)$. In this plane, increasing the cycle-completion rate moves a system upward, reducing residual effective sticking moves it leftward, and reducing the wall-plug-equivalent muon cost or increasing the useful cycle energy relaxes the gain boundary. The same representation therefore separates rate-limited, sticking-limited, and cost-limited regimes, and gives a direct way to compare historical high-yield anchors with proposed improvements such as resonant molecular formation in compressed targets, external-field-assisted reactivation, and higher-intensity or lower-cost muon sources \cite{iiyoshi2019muon,Yamashita:2022rtu,mori2021enforced,kimura2008application,Liu:2022gbs,Cai:2024HighIntensityMuon,Shimomura:2024MUSE,Xu:2025HIAFMuon}. The paper is organized as follows. Section~\ref{sec:lawson_to_mucf} derives the cycle-closure criterion from the standard \mucf{} renewal kinetics and clarifies its relation to the classical Lawson argument. Section~\ref{sec:gain_boundary} introduces the energy and gain boundaries, including the conditional sticking no-go condition. Section~\ref{sec:map} projects representative historical and prospective cases onto the closure map. Section~\ref{sec:discussion_conclusion} discusses the implications for energy-oriented \mucf{} and neutron-source applications.


\section{From Lawson Balance to Muon-Cycle Closure}
\label{sec:lawson_to_mucf}

The classical Lawson criterion is not used here as a variable-by-variable template for \mucf{}. Its role is more limited and more physical: it identifies what a closure criterion must accomplish. In a thermonuclear plasma, a large microscopic fusion cross section is insufficient unless the reacting fuel remains available long enough, at sufficient density and temperature, for the released energy to compensate the losses \cite{Lawson1957Criterion}. A D--T \mucf{} system has a different state variable, a different residence mechanism, and a different dominant loss structure. The relevant object is not a confined thermal plasma element, but the stochastic life history of one negative muon as it cycles through muonic-atom formation, $\dtmu$ formation, fusion, release, possible $\alphamu$ sticking, reactivation, and ordinary muon decay \cite{ponomarev1990muon,froelich1992muon,Petitjean:1992iq}. Therefore the \mucf{} closure criterion must be derived from the renewal balance of this single-muon cycle itself, not from a formal substitution of $n\tau_E T$ by another product. The assumptions made below are correspondingly minimal: the many-cycle history of one useful muon is represented by an effective cycle-completion rate $\Lamc$, ordinary muon decay provides the dominant lifetime scale $\lambda_\mu=1/\taumu$, and all post-fusion alpha-sticking and reactivation physics is summarized by a residual effective sticking probability $\weff$. Additional non-catalytic losses, such as nuclear capture in a bound muonic atom, can be absorbed into an effective loss rate without changing the algebraic form of the reduced criterion.

Under these assumptions, the average fusion yield follows from the renewal balance of the active muon population. While a muon remains available for catalysis, effective cycle completions occur with rate $\Lamc$; each completed cycle produces one D--T fusion event and returns the muon to the active pool with probability $1-\weff$. The same active muon is also removed from the active population by decay, represented here by the dominant rate $\lambda_\mu=1/\taumu$. Hence the total removal rate from the active catalytic population is $\lambda_\mu+\weff\Lamc$, whereas the fusion production rate per active muon is $\Lamc$. The mean number of fusions generated by one useful muon is therefore
\begin{equation}
\Nfus
=
\frac{\Lamc}{\lambda_\mu+\weff\Lamc}.
\label{eq:mucf_yield_rate}
\end{equation}
Introducing the dimensionless cycle strength $\Lmu=\Lamc\taumu$, Eq.~\eqref{eq:mucf_yield_rate} becomes
\begin{equation}
\Nfus
=
\frac{\Lmu}{1+\weff\Lmu}.
\label{eq:mucf_yield_dimensionless}
\end{equation}
Equation~\eqref{eq:mucf_yield_dimensionless} is equivalent to the standard \mucf{} yield estimate $Y_f\simeq(\omega_S^{\rm eff}+\lambda_0/\lambda_c\phi)^{-1}$ when $\Lamc$ is identified with the effective cycle-completion rate $\lambda_c\phi$ and $\lambda_0$ with $\lambda_\mu$ \cite{ponomarev1990muon,froelich1992muon,Petitjean:1992iq,Kamimura:2021msf}. The notation $\Lamc$ is used here to emphasize that the relevant rate is not the intramolecular fusion rate of the already formed $(\dtmu)_{J=v=0}$ molecule, but the effective rate for completing an entire catalytic cycle under specified target conditions.

Equation~\eqref{eq:mucf_yield_dimensionless} also provides the operational coordinates used in the closure map. If a measurement or kinetic analysis gives a fusion yield $Y_f$ together with a reported, inferred, or adopted effective sticking probability $\weff$, the corresponding cycle strength can be inferred as
\begin{equation}
\Lmu(Y_f,\weff)
=
\frac{Y_f}{1-\weff Y_f}.
\label{eq:mucf_inverse_lmu}
\end{equation}
This inversion is meaningful only when
\begin{equation}
\weff Y_f<1,
\label{eq:mucf_kinetic_bound}
\end{equation}
which is the purely kinetic statement that the measured yield has not exceeded the sticking-limited saturation value. In the rate-limited regime, $\weff\Lmu\ll1$, Eq.~\eqref{eq:mucf_yield_dimensionless} reduces to $\Nfus\simeq\Lmu$; in the sticking-limited regime, $\weff\Lmu\gg1$, it approaches $\Nfus\simeq1/\weff$. Thus increasing the effective cycle-completion rate and reducing the residual effective sticking probability act in physically different directions: the former increases $\Lmu$, whereas the latter raises the saturation ceiling itself. The strong sensitivity of the inferred cycle strength near the sticking boundary is quantified by
\begin{equation}
\frac{\partial \Lmu}{\partial \weff}
=
\frac{Y_f^2}{(1-\weff Y_f)^2}.
\label{eq:mucf_lmu_sensitivity}
\end{equation}
Equations~\eqref{eq:mucf_inverse_lmu}--\eqref{eq:mucf_lmu_sensitivity} are the basis for projecting historical D--T \mucf{} yield data, together with reported or inferred effective-sticking values, onto the $(\weff,\Lmu)$ plane and for distinguishing genuine rate improvement from an apparent gain caused by a different effective-sticking assumption \cite{Ackerbauer:1999,Bom:2005PAN,Kamimura:2021msf}.

The two coordinates $\Lmu$ and $\weff$ should be understood as effective cycle-level quantities. The residual sticking probability is not the initial alpha-sticking probability alone. If $\omega_S^0$ denotes the probability that the muon is initially captured into an $\alpha\mu$ bound state after fusion, and $R$ denotes the probability that this bound muon is reactivated during subsequent slowing and collisions, then
\begin{equation}
\weff
=
\omega_S^0(1-R).
\label{eq:effective_sticking_definition}
\end{equation}
Microscopic few-body calculations determine $\omega_S^0$, whereas collisional stripping, target density, target geometry, and possible external-field-assisted reactivation enter through $R$ \cite{Rafelski:1989Reactivation,Kamimura:2021msf,mori2021enforced,kimura2008application,Liu:2022gbs}. Similarly, $\Lamc$ is an effective cycle-completion rate that coarse grains the serial processes of muonic-atom formation, isotope transfer, resonant molecular formation, fusion, slowing, transport, and target-state effects. Treating $\Lmu$ and $\weff$ as coordinates therefore does not assume that they are microscopically independent; it provides a controlled way to display how different physical improvements act on the same cycle-closure problem. The wall-plug-equivalent cost of producing and delivering the muon has not yet entered the argument. It enters only when the yield in Eq.~\eqref{eq:mucf_yield_dimensionless} is compared with a specified gain target, as discussed in the next section.


\section{One-Muon Gain Boundary and Sticking No-Go Condition}
\label{sec:gain_boundary}

The kinetic yield derived above becomes a closure criterion only after an energy accounting convention is specified. Let $\Euse$ denote the useful energy assigned to one completed D--T catalytic cycle. Depending on the purpose of the estimate, $\Euse$ may represent only the bare fusion release, $17.6~{\rm MeV}$, or an effective recoverable value that includes blanket and fuel-cycle credits. We also introduce a lumped system factor $\eta_{\rm sys}$ and an effective wall-plug-equivalent muon cost $\Ecost$, defined as the accounting cost required to produce, capture, transport, moderate, and stop one useful negative muon in the target. The useful energy generated per incident useful muon is then
\begin{equation}
E_{\rm out,\mu}
=
\eta_{\rm sys}\Euse\Nfus .
\label{eq:useful_energy_per_muon}
\end{equation}
The corresponding one-muon gain is
\begin{equation}
\Gmu
=
\frac{E_{\rm out,\mu}}{\Ecost}
=
\frac{\eta_{\rm sys}\Euse}{\Ecost}\Nfus .
\label{eq:one_muon_gain_definition}
\end{equation}
Here $\Gmu$ is not intended to replace a full plant-level engineering gain; it is a local closure variable that compares the useful energy generated by the finite catalytic history of one muon with the effective wall-plug-equivalent cost of making that muon available for \mucf{} \cite{ponomarev1990muon,froelich1992muon,Petitjean:1992iq,Kamimura:2021msf}.

It is useful to separate the energy accounting from the kinetic cycle physics by defining the cycle demand
\begin{equation}
N_L
=
\frac{\Ecost}{\eta_{\rm sys}\Euse}.
\label{eq:cycle_demand}
\end{equation}
For $\Gmu=1$, $N_L$ is the number of completed D--T catalytic cycles that would be required in an ideal loss-free sequence to compensate the wall-plug-equivalent cost of one useful muon. For a general target gain $\Gmu$, the required ideal number of cycles is $\Gmu N_L$. Substituting Eq.~\eqref{eq:mucf_yield_dimensionless} into Eq.~\eqref{eq:one_muon_gain_definition} gives
\begin{equation}
\Gmu
=
\frac{1}{N_L}
\frac{\Lmu}{1+\weff\Lmu}.
\label{eq:gain_lmu}
\end{equation}
Solving Eq.~\eqref{eq:gain_lmu} for the cycle strength required to reach a specified target gain gives
\begin{equation}
\Lmu^{\rm req}
=
\frac{\Gmu N_L}{1-\weff\Gmu N_L}.
\label{eq:required_lmu}
\end{equation}
Equation~\eqref{eq:required_lmu} is finite only if
\begin{equation}
\weff
<
\wcrit
\equiv
\frac{1}{\Gmu N_L}.
\label{eq:sticking_no_go}
\end{equation}
This is the conditional sticking no-go boundary. It is conditional because it is defined only after the gain target, wall-plug-equivalent muon-cost accounting, useful cycle energy, and system factor have been specified. Its physical meaning is direct: if Eq.~\eqref{eq:sticking_no_go} is violated, increasing the effective cycle-completion rate cannot reach the desired gain, because the asymptotic yield at $\Lmu\rightarrow\infty$ is limited by $1/\weff$.

The criterion can now be applied as a three-step test. For a specified accounting choice $(\Ecost,\Euse,\eta_{\rm sys})$ and a target gain $\Gmu$, one first computes $N_L$ from Eq.~\eqref{eq:cycle_demand}. One then checks the sticking boundary in Eq.~\eqref{eq:sticking_no_go}. If the system lies on the forbidden side of this boundary, no finite value of the effective cycle-completion rate $\Lamc$ can reach the target gain within the specified accounting convention. If the sticking condition is satisfied, the required cycle strength is obtained from Eq.~\eqref{eq:required_lmu} and compared with the actual or inferred value of $\Lmu$. This operational test may be summarized as
\begin{equation}
\begin{aligned}
&N_L=\frac{\Ecost}{\eta_{\rm sys}\Euse},\\
&\weff<\frac{1}{\Gmu N_L},\\
&\Lmu\geq
\frac{\Gmu N_L}{1-\weff\Gmu N_L}.
\end{aligned}
\label{eq:operational_closure_test}
\end{equation}
The separation between the second and third lines of Eq.~\eqref{eq:operational_closure_test} is essential. The second line determines whether the system is sticking-limited in principle for the chosen target, whereas the third line determines whether the available cycle-completion rate is large enough once the sticking constraint is not prohibitive. Thus cycle-completion-rate improvement, sticking reduction, and wall-plug-equivalent muon-cost reduction are not interchangeable remedies: they move a system through different parts of the closure problem. The next section visualizes Eq.~\eqref{eq:required_lmu} and Eq.~\eqref{eq:sticking_no_go} in the $(\weff,\Lmu)$ plane.

\section{Cycle-Closure Map and Historical Anchors}
\label{sec:map}

\begin{figure*}[htbp]
\centering
\includegraphics[width=0.96\textwidth]{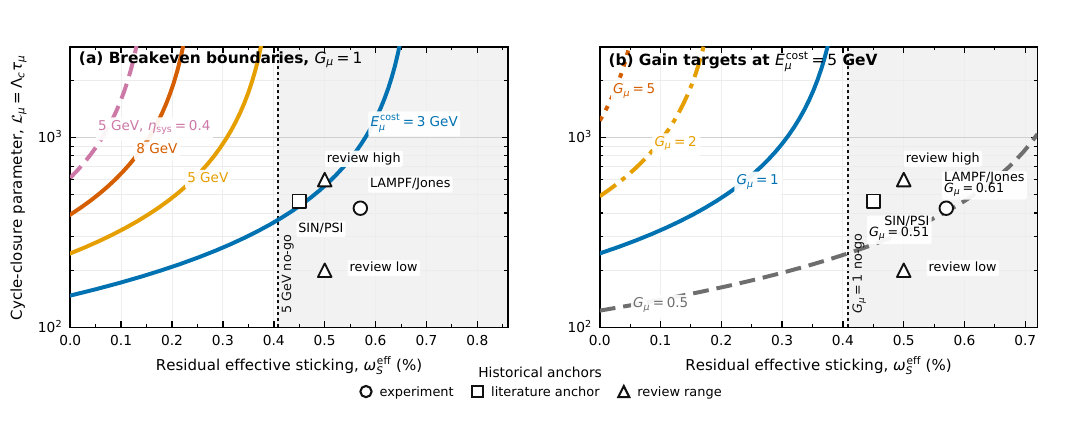}
\caption{
Cycle-closure map for D--T muon-catalyzed fusion in the $(\weff,\Lmu)$ plane.
The horizontal axis is the residual effective sticking probability, displayed in percent; in the equations $\weff$ is dimensionless.
The vertical axis is the cycle-closure parameter $\Lmu=\Lamc\taumu$.
Curves are calculated from Eq.~\eqref{eq:required_lmu}, and the vertical dashed no-go boundaries are given by Eq.~\eqref{eq:sticking_no_go}.
Panel~(a) shows $G_\mu=1$ breakeven boundaries for representative wall-plug-equivalent muon-cost and efficiency assumptions.
Panel~(b) fixes $\Ecost=5~{\rm GeV}$ and shows gain-target boundaries for $G_\mu=0.5$, $1$, $2$, and $5$.
The SIN/Crowe anchor uses the reported low-temperature D--T yield $Y_f=124\pm10$, the effective-sticking value $\weff=0.57\%$, and the reported effective cycle-completion rate $\Lamc=1.93\times10^8~{\rm s^{-1}}$ \cite{Crowe1987LowT}.
The LAMPF/Jones anchor represents the historical high-yield scale of about $150$ fusions per muon \cite{Jones1986Nature,Jones1987Can250}.
The review-range anchors use the Petitjean summary that D--T \mucf{} experiments demonstrated about $100$--$150$ fusions per muon with an effective sticking value of order $0.5\%$ \cite{Petitjean1992OECD}.
}
\label{fig:lawson_map}
\end{figure*}

Figure~\ref{fig:lawson_map} visualizes the closure criterion derived above in the $(\weff,\Lmu)$ plane. The horizontal coordinate is the residual effective sticking probability and is displayed in percent, whereas $\weff$ is used as a dimensionless probability in all equations. The vertical coordinate is the cycle-closure parameter $\Lmu=\Lamc\taumu$, which measures the number of effective cycle-completion opportunities available within one muon lifetime. The curves are not fitted boundaries; they are direct evaluations of Eq.~\eqref{eq:required_lmu}, and the vertical dashed no-go lines are given by Eq.~\eqref{eq:sticking_no_go}. The divergence of each boundary as $\weff$ approaches $\wcrit$ is the analytic sticking limit of Eq.~\eqref{eq:sticking_no_go}, not a numerical artifact. Panel~(a) fixes the target at $G_\mu=1$ and compares representative wall-plug-equivalent muon-cost and efficiency assumptions, including $\Ecost=3$, $5$, and $8~{\rm GeV}$, as well as a more conservative $\Ecost=5~{\rm GeV}$ case with $\eta_{\rm sys}=0.4$. The historical $5~{\rm GeV}$ scale and the corresponding $250$-fusion breakeven estimate follow the conventional accounting discussed by Jones, where approximately $5000~{\rm MeV}$ is assigned as the wall-plug-equivalent cost of one delivered useful negative muon and about $20~{\rm MeV}$ to one useful D--T fusion cycle \cite{Jones1987Can250}. Panel~(b) fixes $\Ecost=5~{\rm GeV}$ and varies the target gain, showing the boundaries for $G_\mu=0.5$, $1$, $2$, and $5$. The overlaid anchors are not fitted to the present model; they are representative projections of reported or review-level D--T \mucf{} yields together with reported, inferred, or adopted effective-sticking values, including the SIN/Crowe low-temperature point, the LAMPF/Jones high-yield anchor, and the Petitjean review range \cite{Crowe1987LowT,Jones1986Nature,Jones1987Can250,Petitjean1992OECD}.

The numerical anchors in Fig.~\ref{fig:lawson_map} are constructed from the parameter set summarized in Table~\ref{tab:historical_projection_inputs}. For the SIN/Crowe point, the reported effective cycle-completion rate gives $\Lmu=\Lamc\taumu$, while the same point is also consistent with the inverse projection in Eq.~\eqref{eq:mucf_inverse_lmu} when $Y_f=124$ and $\weff=0.57\%$ are used. For the LAMPF/Jones anchor, the literature yield $Y_f\simeq150$ is combined with the adopted effective-sticking value $\weff=0.45\%$ to infer $\Lmu$ through Eq.~\eqref{eq:mucf_inverse_lmu}. The Petitjean review anchors are not single experimental measurements; they bracket the review-level range $Y_f=100$--$150$ at an adopted $\weff\simeq0.5\%$. The gain labels associated with the experimental anchors in panel~(b) are evaluated using Eq.~\eqref{eq:one_muon_gain_definition} with the same $\Ecost=5~{\rm GeV}$ wall-plug-equivalent convention used for the plotted gain boundaries. Because Eq.~\eqref{eq:mucf_lmu_sensitivity} shows that the inferred $\Lmu$ becomes highly sensitive when $\weff Y_f$ approaches unity, the plotted points should be read as literature anchors defining the historical parameter region rather than as precision remeasurements.

\begin{table*}[htbp]
\centering
\caption{
Historical and reference inputs used for the cycle-closure projection in Fig.~\ref{fig:lawson_map}.
The effective sticking probability $\weff$ is a reported, inferred, or adopted cycle-level value depending on the anchor; it is listed in percent, but is used as a dimensionless probability in the calculations.
The gain column is evaluated with $\Ecost=5~{\rm GeV}$, $\Euse=20.4~{\rm MeV}$, and $\eta_{\rm sys}=1$.
}
\label{tab:historical_projection_inputs}
\small
\setlength{\tabcolsep}{4.5pt}
\renewcommand{\arraystretch}{1.18}
\begin{tabular}{@{}lllllll@{}}
\toprule
Anchor
& Type
& Input
& $\weff$
& Projection
& $\Lmu$
& $G_\mu$ \\
\midrule
SIN/Crowe
& experiment
& \begin{tabular}[t]{@{}l@{}}
$Y_f=124\pm10$\\
$\Lamc=1.93\times10^8~{\rm s^{-1}}$
\end{tabular}
& $0.57\%$
& $\Lmu=\Lamc\taumu$
& $4.24\times10^2$
& $0.51$
\\
LAMPF/Jones
& literature anchor
& $Y_f\simeq150$
& $0.45\%$
& Eq.~\eqref{eq:mucf_inverse_lmu}
& $4.62\times10^2$
& $0.61$
\\
Petitjean low
& review range
& $Y_f\simeq100$
& $0.50\%$
& Eq.~\eqref{eq:mucf_inverse_lmu}
& $2.00\times10^2$
& $0.41$
\\
Petitjean high
& review range
& $Y_f\simeq150$
& $0.50\%$
& Eq.~\eqref{eq:mucf_inverse_lmu}
& $6.00\times10^2$
& $0.61$
\\
Jones breakeven estimate
& reference target
& $Y_f\simeq250$
& --
& $5000~{\rm MeV}/20~{\rm MeV}$
& --
& $1.0$
\\
\bottomrule
\end{tabular}

\vspace{2mm}
\begin{minipage}{0.96\textwidth}
\footnotesize
The SIN/Crowe point uses the reported low-temperature D--T yield and the reported effective cycle-completion rate \cite{Crowe1987LowT}.
The LAMPF/Jones row is used as a historical high-yield anchor at about $150$ fusions per muon, with $\weff$ treated as an adopted effective-sticking value for projection \cite{Jones1986Nature}.
The Jones breakeven estimate summarizes the conventional wall-plug-equivalent energy-accounting target of roughly $250$ fusions per muon for a $5~{\rm GeV}$ useful-muon cost and about $20~{\rm MeV}$ useful energy per D--T cycle \cite{Jones1987Can250}.
The Petitjean rows represent review-level low and high anchors rather than individual experimental measurements \cite{Petitjean1992OECD}.
\end{minipage}
\end{table*}

Panel~(a) gives the breakeven reading of the map. For $\Euse=20.4~{\rm MeV}$ and $\eta_{\rm sys}=1$, the ideal cycle demands are $N_L\simeq147$, $245$, and $392$ for $\Ecost=3$, $5$, and $8~{\rm GeV}$, respectively. The corresponding sticking boundaries are therefore $\wcrit\simeq0.68\%$, $0.41\%$, and $0.26\%$. With the same $\Ecost=5~{\rm GeV}$ but a more conservative $\eta_{\rm sys}=0.4$, the effective demand increases to $N_L\simeq613$, moving the sticking boundary to $\wcrit\simeq0.16\%$. This shift is the main message of panel~(a): lowering the wall-plug-equivalent muon cost, increasing the useful cycle energy, or improving the system efficiency relaxes the breakeven boundary, whereas a more conservative efficiency assumption rapidly moves the system into the sticking-limited region. The historical anchors lie close to this transition. Under a $5~{\rm GeV}$ wall-plug-equivalent muon-cost convention, effective sticking values in the range $0.45\%$--$0.57\%$ already exceed the $G_\mu=1$ sticking boundary, so increasing the effective cycle-completion rate alone would not close the energy balance. Under a more favorable $3~{\rm GeV}$ wall-plug-equivalent cost convention, the same anchors fall to the allowed side of the sticking boundary, but still require sufficiently large $\Lmu$ to reach breakeven. This is why the distinction between cost-limited, sticking-limited, and rate-limited behavior is visible directly in the left panel.

Panel~(b) gives the gain-target reading at fixed $\Ecost=5~{\rm GeV}$. In this panel the wall-plug-equivalent cost convention is held fixed, so the movement of the curves is caused only by raising the target value of $G_\mu$. The historical anchors fall near the $G_\mu\simeq0.5$--$0.6$ region under the adopted $\Euse=20.4~{\rm MeV}$ and $\eta_{\rm sys}=1$ accounting, consistent with the labels shown for the SIN/Crowe and LAMPF/Jones points. However, the same anchors remain on the wrong side of the $G_\mu=1$ sticking boundary for the $5~{\rm GeV}$ cost convention, and are far from the more demanding $G_\mu=2$ and $G_\mu=5$ contours. This illustrates a useful diagnostic distinction. A point lying below a target contour but to the left of its no-go line is rate-limited: increasing the effective cycle-completion rate can in principle move it upward into the allowed region. A point lying to the right of the no-go line is sticking-limited for that target: reducing $\weff$ is required before any finite cycle-completion-rate improvement can close the gap. A reduction of $\Ecost$, or equivalently an increase of $\eta_{\rm sys}\Euse$, does not move the historical points themselves; it moves the target contours and no-go boundaries. The closure map therefore separates three improvement directions that are often conflated in yield-only discussions: faster cycle completion, lower residual sticking, and lower wall-plug-equivalent useful-muon cost.

The map should therefore be read as a classification tool rather than as a binary verdict on \mucf{}. The historical anchors show that D--T \mucf{} has already reached a high-yield region, but they also show why yield alone is an incomplete metric. Under the conventional $5~{\rm GeV}$ wall-plug-equivalent muon-cost accounting, the path to $G_\mu=1$ requires both movement to the left, through a reduction of $\weff$, and sufficient movement upward, through an increase of the effective cycle-completion strength $\Lmu$. These two directions correspond to different physical programs. External or collisional reactivation primarily reduces the residual sticking probability, whereas improved molecular formation, target compression, transport, and stopping efficiency primarily change the effective cycle-completion rate or the useful-muon cost \cite{Rafelski:1989Reactivation,iiyoshi2019muon,Yamashita:2022rtu,mori2021enforced,kimura2008application,Liu:2022gbs,Cai:2024HighIntensityMuon,Shimomura:2024MUSE,Xu:2025HIAFMuon}. At the same time, sub-breakeven regions of the map are not irrelevant: a system with $G_\mu<1$ may still be valuable as an intense D--T neutron source if the objective is isotope production, irradiation, or transmutation rather than net energy production. The following discussion uses this separation to assess which physical improvement is rate-limited, which is sticking-limited, and which is primarily cost-limited.


\section{Discussion and Outlook}
\label{sec:discussion_conclusion}

The closure map should be understood as a reduced diagnostic criterion, not as a complete reactor model. Its main role is to reorganize the usual yield discussion into three physically distinct quantities: the cycle strength $\Lmu$, the residual effective sticking probability $\weff$, and the energy-demand parameter $N_L=\Ecost/(\eta_{\rm sys}\Euse)$. In this form, a proposed improvement can be classified by asking whether it increases the effective cycle-completion rate, reduces post-fusion muon loss, or lowers the wall-plug-equivalent cost assigned to one delivered useful negative muon. This separation is important because the three effects are not equivalent. Once a system lies on the sticking-forbidden side of Eq.~\eqref{eq:sticking_no_go}, increasing $\Lamc$ alone cannot close the gain balance; conversely, a system below a target contour but still to the left of the no-go line is primarily rate-limited rather than strictly sticking-limited.

The same criterion can be used in reverse to set quantitative design targets. For a specified gain target $\Gmu$ and energy accounting $N_L$, a finite cycle strength $\Lmu$ implies a maximum tolerable residual sticking probability,
\begin{equation}
\omega_{S,\max}^{\rm eff}(\Lmu)
=
\frac{1}{\Gmu N_L}
-
\frac{1}{\Lmu}.
\label{eq:max_sticking_finite_rate}
\end{equation}
This condition is stronger than the asymptotic sticking boundary because it accounts for the actual, finite value of $\Lmu$. It gives a direct benchmark for collisional stripping, density-assisted reactivation, and external-field-assisted reactivation. In particular, cyclotron-resonance stripping, X-ray or laser-assisted stripping, and rate-network descriptions of external-field reactivation should be evaluated by whether they can move $\weff$ below Eq.~\eqref{eq:max_sticking_finite_rate}, rather than by whether they reduce sticking qualitatively \cite{Rafelski:1989Reactivation,kimura2008application,mori2021enforced,Liu:2022gbs,Kou:2026vdz}. Similarly, for a given pair $(\Lmu,\weff)$, the largest wall-plug-equivalent muon cost compatible with a target gain is
\begin{equation}
E_{\mu,\max}^{\rm cost}
=
\frac{\eta_{\rm sys}\Euse}{\Gmu}
\frac{\Lmu}{1+\weff\Lmu}.
\label{eq:max_muon_cost}
\end{equation}
Equation~\eqref{eq:max_muon_cost} converts the statement that the wall-plug-equivalent useful-muon cost must be reduced into a quantitative requirement on muon production, capture, transport, moderation, and stopping efficiency. This is the natural place where accelerator-based muon-source studies, including MUSE-type beamlines, HIAF-based GeV muon concepts, and high-intensity source proposals, enter the closure criterion \cite{Shimomura:2024MUSE,Cai:2024HighIntensityMuon,Xu:2025HIAFMuon}.

These inverse forms clarify how different research directions should be judged. Improvements in resonant molecular formation, in-flight \mucf{}, high-density target platforms, target compression, muon slowing, and target stopping efficiency mainly act by increasing $\Lmu$ or reducing the wall-plug-equivalent useful-muon cost \cite{iiyoshi2019muon,Yamashita:2022rtu,yamashita2025radiative,toyama2026direct,Koukina:2026wlo,Kamimura:2021msf,Wu:2024uad}. Reactivation mechanisms mainly act by reducing $\weff$ \cite{Rafelski:1989Reactivation,mori2021enforced,kimura2008application,Liu:2022gbs,Kou:2026vdz}. Better muon-source performance mainly lowers $N_L$ by reducing the wall-plug-equivalent cost $\Ecost$, but it does not move a historical or experimental point in the $(\weff,\Lmu)$ plane; instead, it moves the gain boundaries. Therefore, a convincing path toward energy-oriented \mucf{} must specify how far the system moves in each coordinate. A proposal that improves only one coordinate may still be valuable, but the closure map shows immediately whether that improvement addresses the active bottleneck.

The same framework is not restricted to net-energy discussions. For neutron-source, isotope-production, irradiation, or transmutation applications, the relevant figure of merit need not be $G_\mu\geq1$ in the thermal-energy sense. One may replace $\Euse$ by an application-dependent neutron utility $\mathcal{U}_X$ per D--T cycle, understood either as an energy-equivalent value or as a normalized application value, and define
\begin{equation}
G_\mu^{(X)}
=
\frac{\mathcal{U}_X \Nfus}{\Ecost}.
\label{eq:utility_gain_generalized}
\end{equation}
Here $X$ may denote isotope production, material irradiation, or waste transmutation. This replacement leaves the kinetic coordinates $\Lmu$ and $\weff$ unchanged, but changes the value assigned to each catalyzed fusion event. Recent isotope-production estimates based on \mucf{} neutrons illustrate why this distinction matters: a system that remains below the energy breakeven boundary may still be useful if the neutron product has sufficiently high application value \cite{Parisi:2025wpj}. Thus the sub-breakeven region in Fig.~\ref{fig:lawson_map} should not be dismissed; it is outside the net-energy target, but not necessarily outside the useful-neutron-source domain.

Several limitations should be kept explicit. The parameters $\Lamc$ and $\weff$ are effective cycle-level quantities and are not microscopically independent. Target density, temperature, molecular formation, muon slowing, collisional stripping, external-field overlap, and post-stripping transport can be strongly coupled. The historical points used above are literature anchors rather than a global refit of all \mucf{} kinetic data. The criterion developed here is therefore best viewed as a common coordinate system: it does not by itself establish feasibility, but it makes explicit whether a proposed \mucf{} system is cycle-completion-rate-limited, sticking-limited, or wall-plug-cost-limited. Future work should couple this reduced criterion to full kinetic networks, energy-dependent muon transport, target geometry, and microscopic reactivation probabilities so that proposed improvements can be placed quantitatively on the same closure map \cite{Kou:2026vdz}.

\begin{acknowledgments}
This work has been supported by the Research Program of State Key Laboratory of Heavy Ion Science and Technology, Institute of Modern Physics, Chinese Academy of Sciences (Grant No. HIST2025CS08). The author gratefully acknowledges the stimulating discussions with colleagues and invited experts at the First Small Workshop on Muon-Catalyzed Fusion, held in June 2026 at the Southern Center for Nuclear-Science Theory, Institute of Modern Physics, Chinese Academy of Sciences, Huizhou, Guangdong, China.
\end{acknowledgments}

\bibliographystyle{apsrev4-2}
\bibliography{refs}

\end{document}